\documentclass[twocolumn]{emulateapj}
\submitted{Science with Parkes @ 50 Years Young, 31 Oct. -- 4 Nov., 2011}

\shorttitle{The Double Pulsar in its 8th anniversary}
\shortauthors{M. Burgay}

\begin{document}

\title{The Double Pulsar System in its 8th anniversary}

\author{Marta Burgay}
\affil{INAF - Osservatorio Astronomico di Cagliari, Loc. Poggio dei
  Pini, Strada 54, 09012 Capoterra (CA), Italy}
\email{burgay@oa-cagliari.inaf.it}

\begin{abstract}
The double pulsar system J0737$-$3039A/B, discovered with the Parkes
radio telescope in 2003, is one of the most intriguing pulsar findings
of the last decade. This binary system, with an orbital period of only
2.4-hr and with the simultaneous presence of two radio pulsed signals,
provides a truly unique laboratory for relativistic gravity and plasma
physics. Moreover its discovery enhances of almost an order of
magnitude the estimate of the merger rate of double neutron stars
systems, opening new possibilities for the current generation of
gravitational wave detectors. In this contribution we summarise the
present results and look at the prospects of future observations.
\end{abstract}


\section{Introduction}

The 22.7-ms binary pulsar PSR J0737-3039A (hereafter 'A') was
discovered in April 2003 \citep{bdp+03} in the data of the Parkes
High-Latitude Pulsar Survey \citep{bjd+06}. Its short orbital period
($P_b = 2.4$ hrs), together with a remarkably high value of the
periastron advance ($\dot\omega = 16.9$ deg/yr) identified it soon as
a member of the most extreme relativistic binary system ever
discovered. The compactness of the system, combined with its short
coalescence time ($T_{coal} = 85$ Myr) and low luminosity, increases
the estimates on the double neutron star coalescence rate by almost an
order of magnitude \citep{bdp+03, kkl+04, kkl+04a}, boosting hopes to
detect mergers of neutron stars (NSs) with ground based gravitational
wave detectors.

Analysis of follow-up observations led, in October 2003 (almost
exactly 8 years before the symposium for the Parkes radio telescope
50th anniversary) to the discovery of a second pulsar in the system
\citep{lbk+04}, the 2.8-s pulsar J0737-3039B (hereafter 'B'). The
reason why the signal of pulsar B was not detected earlier is that
this object is only bright in two short sections of the orbit; for the
rest of the orbit the signal is very week or absent.

Upon closer inspection, the signals of both pulsar A and B revealed
other intriguing characteristics: pulsar A is eclipsed for $\sim 30$ s
near superior conjunction and pulsar B shows variations in the pulse
shape along the orbit \citep{lbk+04}. Variations of the pulse shape on
longer time scales and of the extent and location of B's bright
phases have also been observed \citep{bpm+05, pmk+10}. These phenomena are
likely related to the geodetic precession of pulsar A and B that are
changing the geometry of the system and hence our view towards
it. This relativistic effect happens on such a short time scale on
this system (75 years for pulsar B and 71 for pulsar A) that in 2008
pulsar B's beam went out of sight \citep{pmk+10}.

In this contributions, we will describe the binary system
J0737-3039A/B focusing mainly on its current and future applications
as a test-bed for relativistic theories. 

\section{Testing Gravity Theories with the Double Pulsar}

Because of their strong gravitational fields and rapid orbital
motions, the binary systems containing two neutron stars can exhibit
large relativistic effects \citep{dd86}. If one (or two, as in this
unique case) of the NSs emits clock-like radio pulsed signals it is
possible, by measuring the delays in the time of arrival of the
pulses, to measure directly not only the Keplerian parameters of the
orbit, but also the relativistic corrections to the Keplerian
description of an orbit, the so called ``post-Keplerian'' (PK)
parameters, hence testing relativistic gravity. In each theory of
gravity the PK parameters can be written as a function of
the masses of the two stars and of the measurable Keplerian
parameters. With the two masses as the only unknowns, the measurement
of three or more PK parameters over-constrains the system hence
providing tests for a given theory of gravity \citep{dt92}.

In General Relativity (GR) the post-Keplerian parameters can be
written (at first post-Newtonian order, 1PN) as follows \citep{dd86}:
\begin{small}
\begin{eqnarray*}
\dot{\omega} &=& 3 T_\odot^{2/3} \; \left( \frac{P_b}{2\pi} \right)^{-5/3} \;
               \frac{1}{1-e^2} \; (M_A + M_B)^{2/3}, \label{omegadot}\\
\gamma  &=& T_\odot^{2/3}  \; \left( \frac{P_b}{2\pi} \right)^{1/3} \;
              e\frac{M_B(M_A+2M_B)}{(M_A+M_B)^{4/3}}, \\
\dot{P}_b &=& -\frac{192\pi}{5} T_\odot^{5/3}  \left( \frac{P_b}{2\pi} \right)^{
-5/3}
               \frac{\left(1 +\frac{73}{24}e^2 + \frac{37}{96}e^4 \right)}{(1-e^
2)^{7/2}}
               \frac{M_AM_B}{(M_A + M_B)^{1/3}}, \\
r &=& T_\odot M_B, \\
s &=& T_\odot^{-1/3} \; \left( \frac{P_b}{2\pi} \right)^{-2/3} \; x \;
              \frac{(M_A+M_B)^{2/3}}{M_B},
\end{eqnarray*}
\end{small}

\begin{figure*}
\begin{center}
\includegraphics[width=115mm, clip=true]{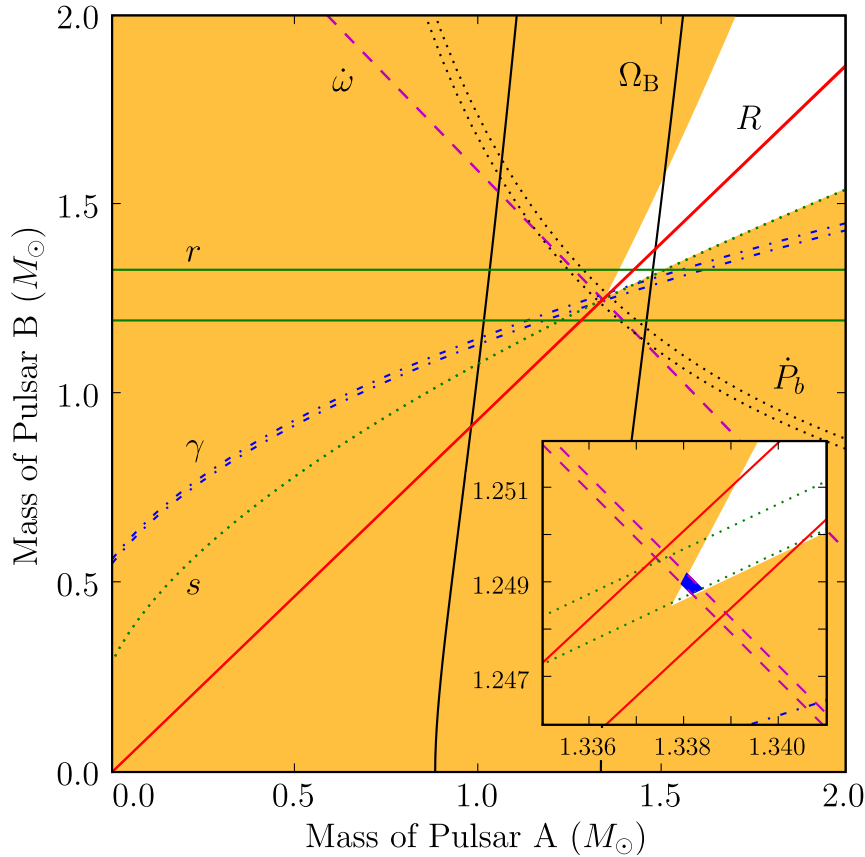}
\caption{The observational constraints on the masses of the two
  pulsars in the double pulsar system, M$_A$ and M$_B$. The coloured
  regions are those which are excluded by the Keplerian mass functions
  of the two pulsars. Further constraints are shown as pairs of lines
  enclosing permitted regions as predicted by general relativity: (a)
  the measurement of the advance of periastron $\dot{\omega}$ (dashed
  lines); (b) the measurement of the mass ratio R (solid lines); (c)
  the measurement of the gravitational red-shift/time dilation
  parameter $\gamma$ (dot-dash lines); (d) the measurement of Shapiro
  parameter $r$ (solid horizontal lines) and Shapiro parameter $s$
  (dotted lines); (e) the measurement of the orbital decay (dotted
  lines); (f) the measurement of the relativistic spin precession
  (solid dark lines). Inset is an enlarged view of the small square
  which encompasses the intersection of the three tightest
  constraints. The permitted regions are those between the pairs of
  parallel lines and we see that an area exists which is compatible
  with all constraints. From \cite{bkk+08}.}
\label{fig:mAmB}
\end{center}
\end{figure*}

where $P_b$ is the orbital period, $e$ the eccentricity and $x$ the
projected semi-major axis of the orbit measured in light-s. The masses
$M_A$ and $M_B$ of A and B respectively (or, in general, of the pulsar
and its companion), are expressed in solar masses. The constant
$T_\odot$ is defined as $T_\odot=GM_\odot/c^3=4.925490947 \mu$s where
$G$ is the Newtonian constant of gravity and $c$ the speed of
light. The first PK parameter, $\dot{\omega}$, describes the
relativistic advance of periastron. The parameter $\gamma$ denotes the
amplitude of delays in arrival times caused by the varying effects of
the gravitational redshift and time dilation as the pulsar moves in
its elliptical orbit at varying distances from the companion and with
varying speeds. The decay of the orbit due to gravitational wave
damping is expressed by the change in orbital period, $\dot{P}_b$. The
other two parameters, $r$ (rate) and $s$ (shape), are related to the
Shapiro delay caused by the gravitational field of the companion.

The PK parameter can be plotted on a mass-mas diagram (see e.g. Fig.
\ref{fig:mAmB}) and, if the theory tested is correct, the curves on
the plane must intersect in a single point.

\begin{figure*}
\begin{center}
\resizebox{\hsize}{!}{\includegraphics[width=7truecm]{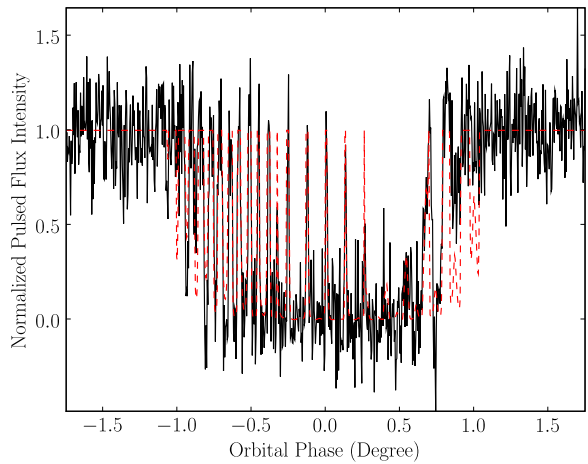},
\includegraphics[width=7truecm]{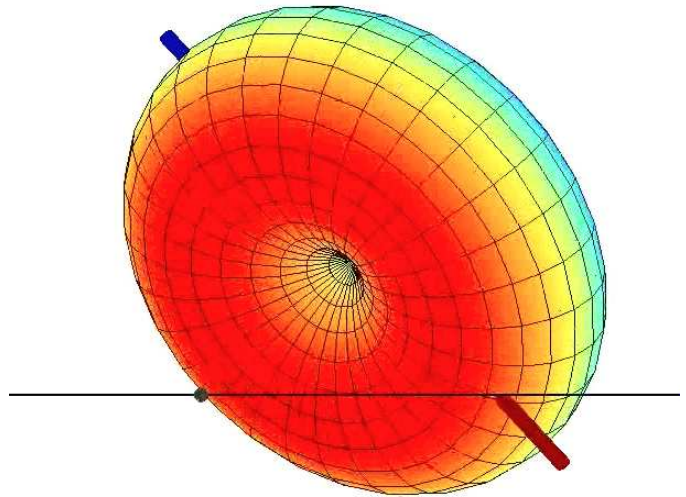}}
\caption{Left: average eclipse profile of pulsar A at 820\,MHz over a
five-day period around April 11, 2007 (black solid line) along with a
model eclipse profile (red dashed line). Near orbital phase 0.0 the
spikes are separated by the spin period of pulsar B while at eclipse
ingress and egress at twice B's period \citep{bkk+08}. Right: geometric model
for B's magnetosphere eclipsing A (black circle) along its orbit
(horizontal line).}
\label{fig:ecli}
\end{center}
\end{figure*}

In this context, PSR J0737-3039A/B, with its two clock-like signal and
its relativistic orbital motion, promises to be the most powerful
instrument to test GR (and other theories). Measurements of the times
of arrival ({\it{timing}}) of pulsar A, in fact, have provided all 5
post-Keplerian parameters with high accuracy after only 3 years of
observations and precision on their measurement increases with time as
we continue monitoring this system. Moreover, with the knowledge of
the projected semimajor-axes for both A and B, possible only in this
system, we obtain a precise measurement of the mass ratio $R$ of the
two stars:
\begin{equation}
R\equiv M_A/M_B=x_B/x_A
\end{equation}
For every realistic theory of gravity, we can expect the mass ratio,
$R$, to follow this simple relation \citep{dt92}, at least to 1PN
order. The $R$ value is not only theory-independent,
but also independent of strong-field (self-field) effects which is not
the case for PK-parameters. This provides a stringent and new
constraint for tests of gravitational theories as any combination of
masses derived from the PK-parameters {\it must} be consistent with
the mass ratio. 

Another effect predicted by relativistic theories is the relativistic
spin precession which, for General Relativity, can be written (for
pulsar A, for instance) as:

\begin{equation}
\Omega_s = \left( \frac{2\pi}{P_b}\right)^{5/3} T_\odot^{2/3}\; \frac{M_B(4M_A+3M_B)}{2(M_A+M_B)^{4/3}}\frac{1}{1-e^2}
\end{equation}

For the two pulsars in the double pulsar system this implies a period
for the relativistic spin precession of 75 years for A and 71 years
for B. Given these small numbers one would expect to see a rapid
variability in the observed pulse profiles caused by the change in
the line of sight through the radio beam as the latter precesses
around the total angular momentum vector (as observed e.g. in pulsar
B1534+12 \citep{atw99, stta00}. In the case of pulsar A no such changes
have been seen in the past 8 years \citep{mkp+05} suggesting that the
spin axis has a very small misalignment with respect to the total
angular momentum vector \citep{fsk+08}. Pulsar B, on the contrary shows
variations on the shape of pulse profile and on the extent of the
bright visibility phases over different time scales, at rates
comparable with that of the relativistic spin precession \citep{bpm+05}
and in 2008 its radio beam has gone outside our line of sight
\citep{pmk+10}. With B's pulse profile variation, anyway, only a
qualitative assessment of the effects of the precession can be
done. The double pulsar system, however, shows another very peculiar
characteristic: pulsar A is eclipsed for about 30 s around superior
conjunction and the light curve in the eclipsed region shows a clear
modulation with the spin period of pulsar B (or twice the period,
depending on the exact orbital phases; see Fig.
\ref{fig:ecli}). Describing this eclipse of pulsar A as due to
absorption occurring in the magnetosphere of pulsar B, \citet{bkk+08}
successfully use a simple geometric model to characterize the observed
changing of the eclipse morphology with time and hence to measure the
relativistic precession of pulsar B.

With five PK parameters already available, and the mass ratio
measurement, this additional constraint makes the double pulsar the
most overdetermined system to date providing five possible tests for
relativistic theories, the most stringent of which, given by the
measurements of $R$, $\dot{\omega}$ and $s$, tests General relativity
at the 0.02\% level (see Fig. \ref{fig:mAmB}.

\begin{figure}
\begin{center}
\resizebox{\hsize}{!}{\includegraphics{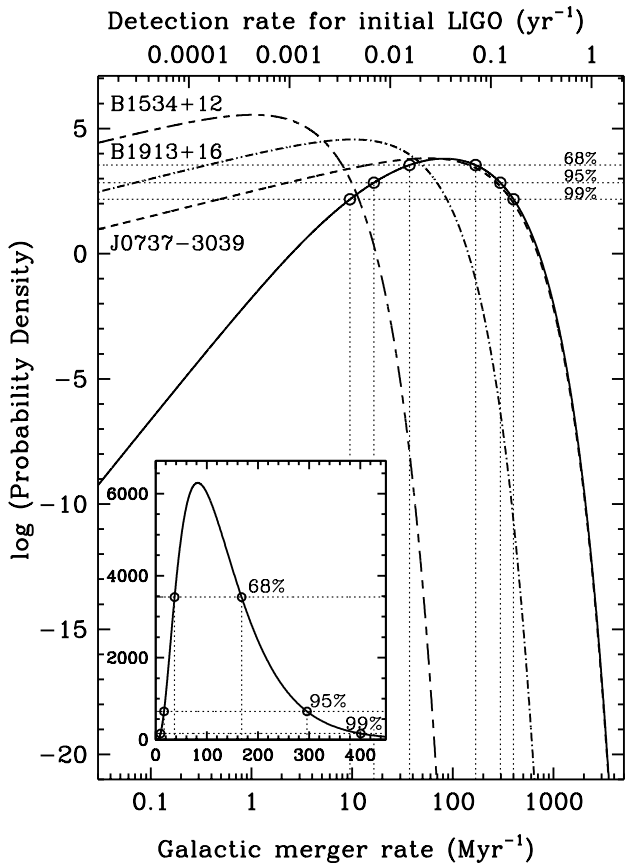}}
\caption{Probability density function representing the expectation
that the actual DNS binary merger rate in the Galaxy (bottom axis) and
the predicted initial LIGO detector rate (top axis) take on particular
values, given the observations. The curves shown are calculated
assuming the parameters of the reference model in \citet{kkl+04}. The
solid line shows the total probability density, along with those
obtained for each of the three known binary systems (dashed lines)
coalescing within a Hubble time. Inset: Total probability density, and
corresponding 68\%, 95\%, and 99\% confidence limits, shown in a
linear scale. From \citet{kkl+04a}}
\label{fig:merger}
\end{center}
\end{figure}

\begin{figure}
\begin{center}
\resizebox{\hsize}{!}{\includegraphics[height=3cm]{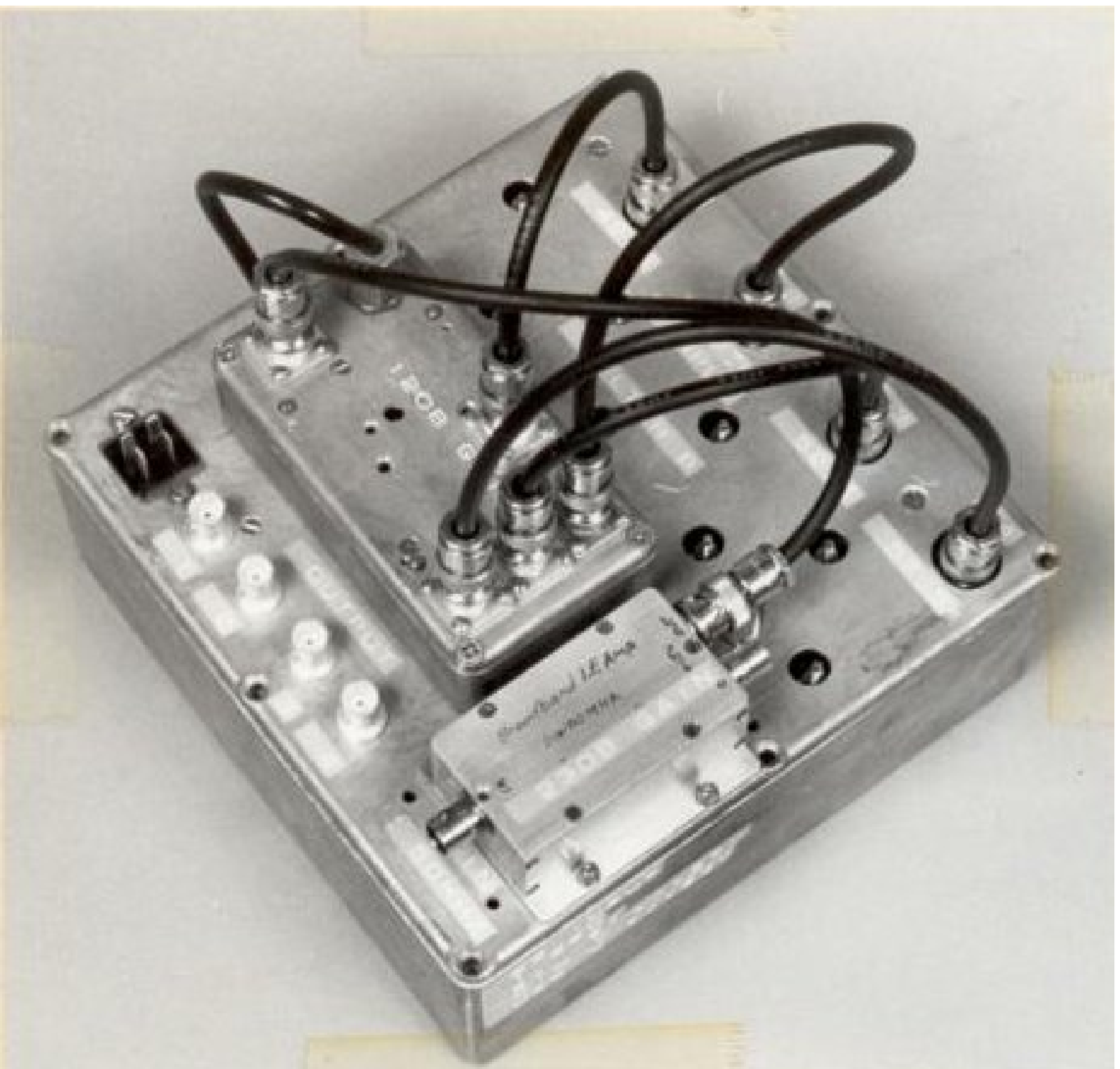}\includegraphics[height=3cm]{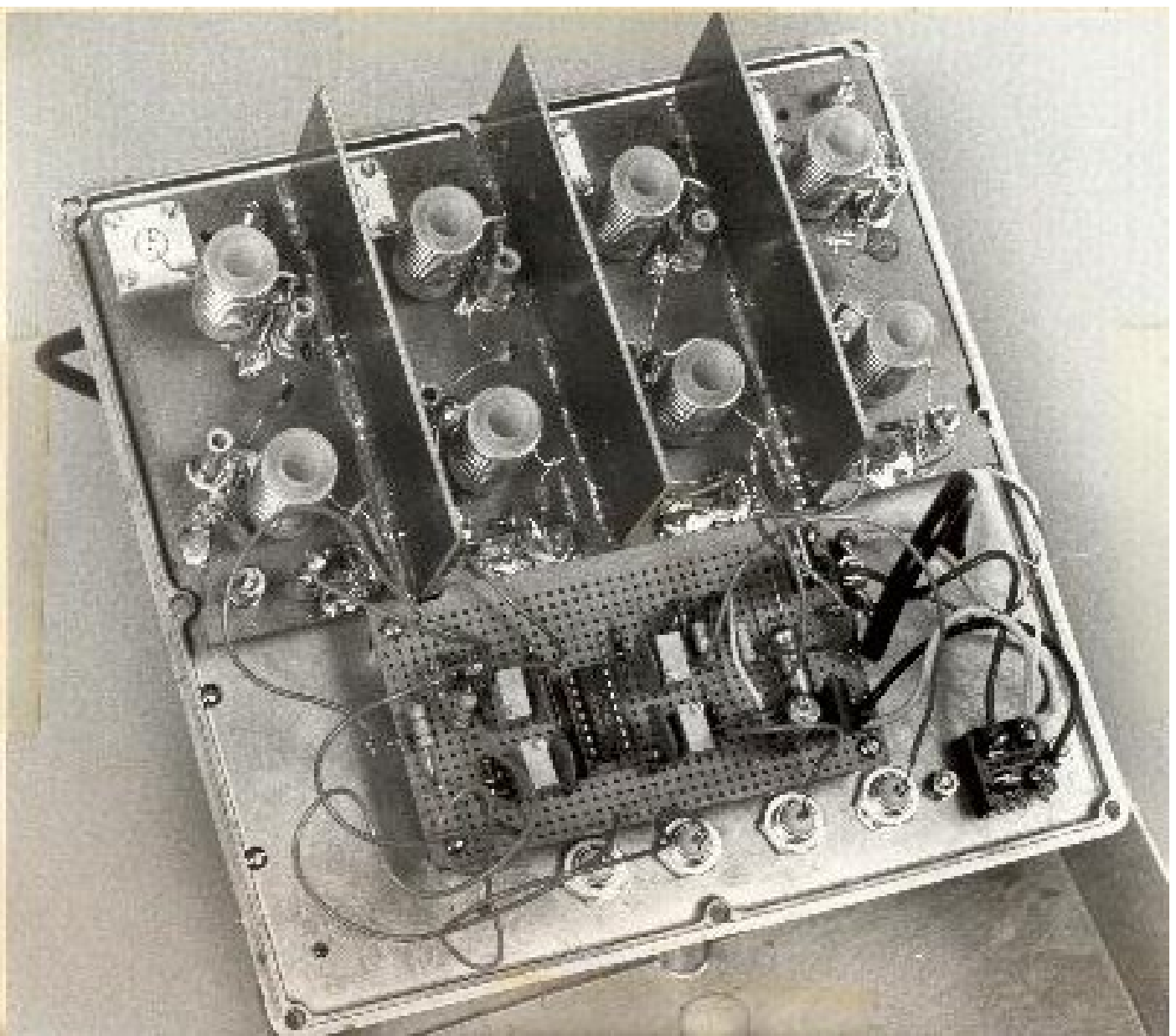}}
\caption{The hand-made filterbank built for the Sample4 experiment.}
\label{fig:sample4}
\end{center}
\end{figure}

\begin{figure}
\begin{center}
\resizebox{\hsize}{!}{\includegraphics{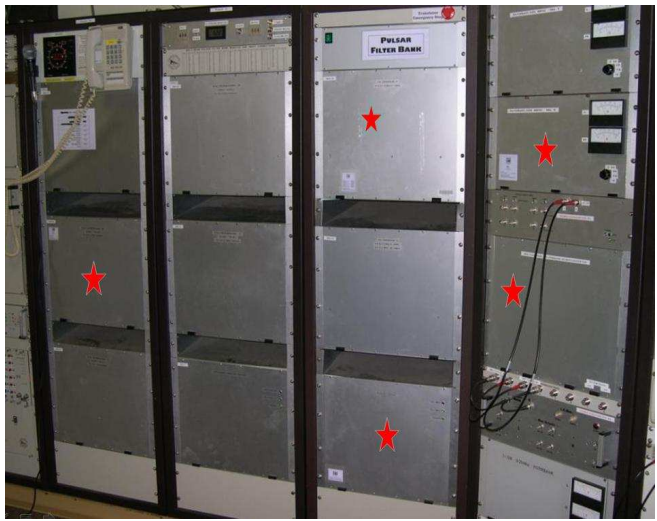}}
\caption{Some of the racks of the Parkes old control room. The red
stars mark various pieces of hardware, including parts of the 21-cm
analogue filterbank, built at the Bologna Astronomical Observatory and
the IRA-Medicina radio telscope (Italy).}
\label{fig:racks}
\end{center}
\end{figure}

\begin{figure*}
\begin{center}
\centering{\includegraphics[width=120mm]{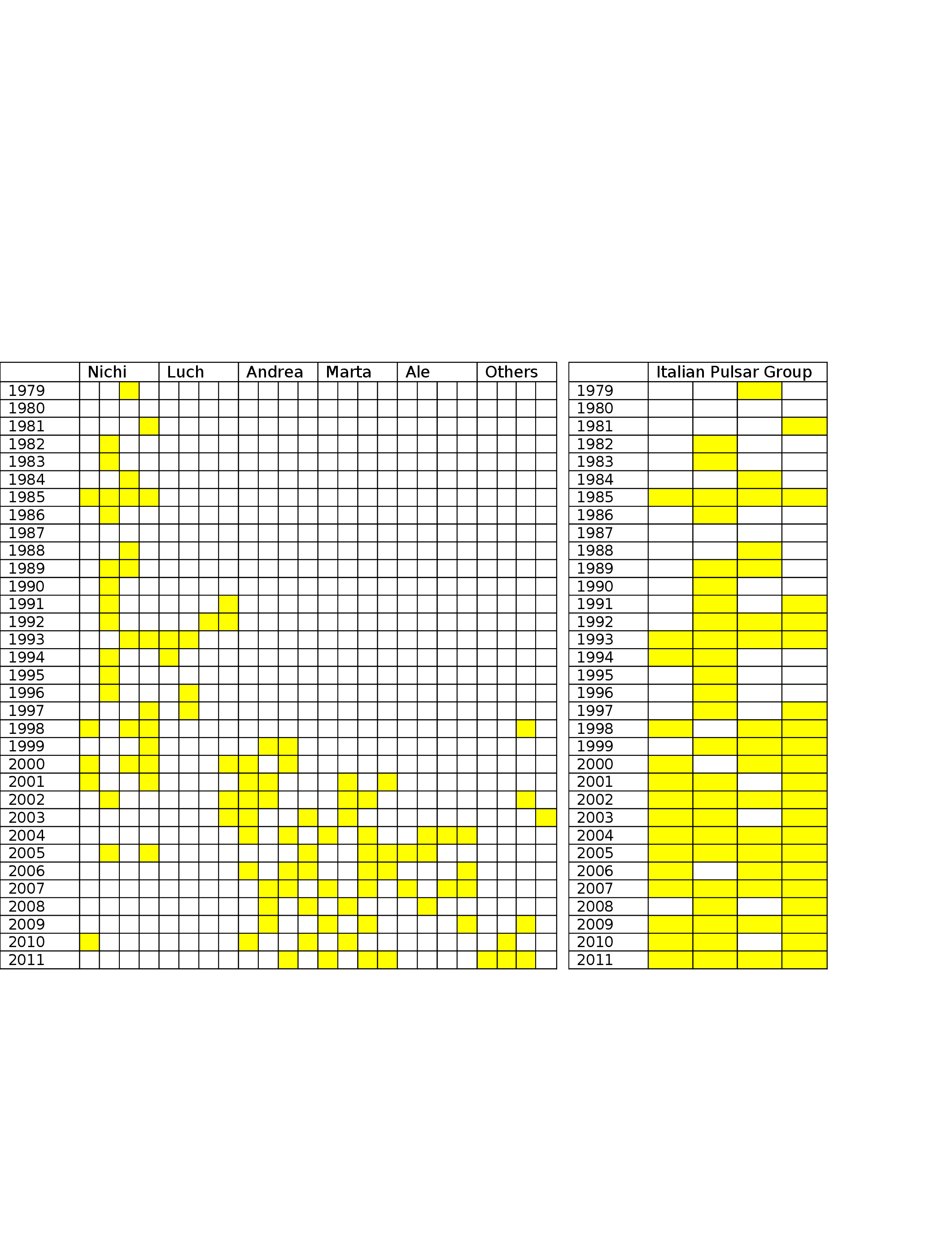}}
\caption{Schematic view of the italian presence at the Parkes
Observatory. In the left part of the grid, a yelow square marks a
visit of a member of the group (whose name is specified in the first
row) during a given quadrimester. On the right, the overrall presence
of all members of Prof. D’Amico’s team at Parkes is summarised (we
never leave the telescope alone!).}
\label{fig:table}
\end{center}
\end{figure*}

\section{An new estimate of double neutron stars coalescence rate}

As mentioned above, for the double pulsar it has been possible to
measure all 5 PK parameters in just 3 years of timing
observations. The last measured, and one of the most interesting for
its implications, is the decay of the orbit $\dot{P}_b$ which,
according to General relativity, is due to emission of gravitational
waves. At the measured rate of $-1.24\times 10^{-12}$, the orbit
should shrink by 7 mm per day, implying that the NSs in the double
pulsar system will coalesce in only 85 Myr. Such a merger event should
produce a burst of gravitational waves detectable from current
generation ground based interferometers. 

A reliable estimate of the double neutron star (DNS) merger rate in
the Galaxy is crucial in order to predict whether these gravity wave
detectors will be successful in detecting such bursts in a reasonable
timescale. Before the discovery of the Double pulsar the estimates of
this rate were rather low \citep{cl95, acw99, knst01, kkl03}, because
only a few DNss with merger times less than the age of the Universe
were known. With the discovery of this system, with a lifetime two
times shorter than the shortest known before (that of PSR B1913+19), a
radio luminosity 6 times smaller and a very short orbital period,
making it more difficult to detect in a blind survey, the estimate
rates have increased by a factor of ~6 \citep{bdp+03, kkl+04, kkl+04a}.

Thanks to the discovery of the double pulsar, hence, the estimated
merger rates and, by consequence, the detection rates for ground based
gravitational wave detectors, fall for the frst time in a human time
scale.

\section{Future experiments}

The double pulsar system J0737-3039A/B has already proven to be an
excellent laboratory for relativistic gravity providing new and more
stringent tests for General Relativity and boosting hopes for the
gravitational wave community. The future promises to be even brighter.
Thanks to the fact that the precision with which the measured
parameters (rotational, positional, orbital - classic and relativistic)
increases with time (see e.g. \citet{kw09}) as we proceed with the
timing observations, we will soon need to include the second
post-newtonian order (2PN) in some of the equation adopted. At 2PN the
expression for General Relativity of the periastron advance
$\dot{\omega}$ depends on the geometry of the system, constrained,
for pulsar A, by the lack of changes in the pulse shape and for B by
the shape of the light curve of A during eclipse, and on the moment of
inertia $I$ of the NSs \citep{ds88}. Measuring $\dot{\omega}$ at 2PN
should allow us to measure $I$ hence to tightly constrain the equation
of state of a NS \citep{kw09}.

Another intriguing possibility is to constrain not only GR but also
alternative theories of gravity: \citet{esp04} shows, for instance,
that a measurement with a 1\% precision of the orbital decay in the
double pulsar system would constrain the $\alpha_0$ and $\beta_0$
parameters for scalar tensor theories better than the current solar
system tests. Also, \citet{wk07} show that measuring a correlation
in the change of the eccentricity and the periastron advance in a
relativistic binary pulsar should be the signature of the existence of
a preferred reference frame with respect to which the orbit is varying
in time.

\section{The Italian Pulsar Connection}

In this final section, on behalf of Nichi D’Amico and the rest of the
Italian pulsar group, I’d like to briefly summarize the 30-years long
story of the ’Italian-pulsar connection’ that tightly links us to the
Parkes radio tele- scope. After having visited Parkes for the first
time in 1979, Nichi started a collaboration with Dick Manchester in 1981, by
building a prototype filter bank (Fig.~\ref{fig:sample4}) designed for pulsar
searches a 21cm. The experiment was called ``Sample4'', as 4 frequency
channels only were sampled at 21cm in order to remove dispersion and
interstel- lar scattering. That experiment represents the first
attempt to ex- ploit relatively high frequencies to search pulsars
hidden in the most dense and dispersed regions of the Galaxy, and put
the basis for a boom of pulsar discoveries which was achieved at
Parkes in the following years. Indeed, since then, larger and more
ambitious experiments were continuously setup at Parkes by the
Italians, the British, and the other international collaborators (Fig.~\ref{fig:racks}), in a fruitful collaboration with Parkes staff, which ultimately
led to the “multibeam” surveys and the discovery of the Double Pulsar
and many other interesting systems. Nichi managed to get many Italian
students and post-docs to get involved in these experiments. An
overall figure of more than 100 visits to Australia by the Italian
pulsar group was registered in the last 30 years (Fig.~\ref{fig:table}), which
represents for all of us an unforgettable testimony of the
professional skills and the human passion of Parkes staff.

\acknowledgments 

Thanks to the Parkes Radio Telescope and its wonderful staff. In these
last 10 years I spent quite a bit of time down-under observing (not
only the double pulsar): I learnt a lot and I enjoyed every moment
(even the 4 AM shifts)!

\end{document}